\documentclass[final,twocolumn]{revtex4}

\usepackage{epsfig,graphicx}
\usepackage{amssymb}
\usepackage{amsfonts}
\usepackage{bm}

\begin{document}

 \title{Di-electron and two-photon widths in charmonium}

 \author{\firstname{A.M.}~\surname{Badalian}}
 \email{badalian@itep.ru} \affiliation{Institute of Theoretical and
 Experimental Physics, Moscow, Russia}

 \author{\firstname{I.V.}~\surname{Danilkin}}
 \email{danilkin@itep.ru} \affiliation{Moscow Engineering Physics Institute, Moscow,
 Russia}\affiliation{Institute of Theoretical and Experimental
 Physics, Moscow, Russia}


 \begin{abstract}
The vector and pseudoscalar decay constants are calculated in the
framework of the Field Correlator Method. Di-electron widths:
$\Gamma_{ee}(J/\psi)=5.41$ keV, $\Gamma_{ee}(\psi'(3686))=2.47$
keV, $\Gamma_{ee}(\psi''(3770))=0.248$ keV, in good agreement with
experiment, are obtained with the same coupling, $\alpha_s=0.165$,
in  QCD radiative corrections. We show that the larger
$\alpha_s=0.191\pm 0.004$ is needed to reach agreement with
experiment for $\Gamma_{\gamma\gamma}(\eta_c)=7.22$ keV,
$\Gamma_{\gamma\gamma} (\chi(^3P_0))=3.3$ keV,
$\Gamma_{\gamma\gamma}(\chi(^3P_2))= 0.54$ keV, and also for
$\Gamma(J/\psi\rightarrow 3g)=59.5$ keV,
 $\Gamma(J/\psi\rightarrow \gamma 2g)=5.7$ keV. Meanwhile even
larger $\alpha_s=0.238$ gives rise  to  good description of
$\Gamma(\psi'\rightarrow 3g)=52.7$ keV,  $\Gamma(\psi'\rightarrow
\gamma 2g)= 3.5$ keV, and provides  correct ratio  of the
branching fractions: $\frac{\mathcal{B}(J/\psi\rightarrow
\textrm{light\ hadrons})}{\mathcal{B}(\psi'\rightarrow
\textrm{light\ hadrons})}=0.24.$
\end{abstract}

 \maketitle

\section{Introduction}
\quad\quad Low-lying states of heavy quarkonia  have been an
important laboratory to study both perturbative and
nonperturbative phenomena in QCD. However, recent discoveries of
higher resonances, in particular  $X(3872)$ \cite{1} and $Y(4260)$
\cite{2} have shown that these and some other new resonances
cannot be interpreted as conventional $Q \bar{Q}$ mesons. To
understand the nature of new resonances, evidently, two- (or
many-) channel consideration  is needed. However, in strict sense
it cannot be done now because nonperturbative theory of strong
decays is not still well developed in QCD. Therefore for
identification of new resonances with $J^{PC}=1^{--}$, observed in
$e^+e^-$ via the initial state radiation  \cite{3}, \cite{4},
 a special role belongs to di-electron widths and  also  two-photon
widths for  C-even resonances, which are reasonably well
described  by existing QCD formulas. At this point it is worthwhile
to remind that   the di-electron width of a $Q \bar{Q}$ meson is
by two orders (may be even more) larger than di-electron width of
a compact four-quark system \cite{5}.

In our paper, firstly, we calculate the decay constants of vector
(V) and pseudoscalar (P) mesons in charmonium using the Field
Correlator Method (FCM), which has been successfully applied to
heavy-light mesons \cite{6}. Due to relativistic corrections
di-electon widths and their ratios, calculated here, agree with
experiment with high accuracy. Therefore from the absolute values
of di-electron widths some important factors, containing the
squared wave functions at the origin,  can be extracted  and then
used in different annihilation decays.

We pay a special attention to the influence of radiative
corrections on different annihilation rates. The absolute values
of  di-electron widths are shown to agree with experimental
numbers only if the QCD radiative corrections are taken into
account. Unfortunately, at present there is no  consensus about
the true value of the strong coupling in them.  These corrections,
known in first (one-loop) approximation
 \cite{7}, \cite{8},  enter the
di-electron and two-photon widths as  separate factors:
$\beta_V=1-\frac{16 \alpha_s}{3\pi}, \beta_P=1-\frac{3.7 \alpha_s}
{\pi}.$  In  \cite{9}  these factors are put equal unity:
$\beta_V=\beta_P=1.0$, i.e. the QCD correction is neglected, while
in \cite{10}-\cite{13} their values are almost two times smaller,
$\beta_V=0.52\pm 0.06$. The reason of this uncertainty partly
occurs because the contribution of  higher corrections  remains unknown.
Therefore, although by derivation the coupling $\alpha_s$ in
different annihilation widths  is defined at the standard scale
$\mu=2m_Q$ or $\mu=M_{V(P)}$ \cite{14} (in the $\overline{MS}$
scheme), factually, this strong coupling appears to be an
effective one and can differ in different annihilation decays,
since  for them higher order perturbative corrections can  be
different.

In our paper we show that in the $\psi$- family the di-electron
widths are described with the same  coupling, which turns out to
be relatively small: $\alpha_s=0.165$ or $\beta_V=0.72$ (the same
"universality" is observed in bottomonium \cite{15}). Meanwhile to
describe two-photon widths of $\eta_c$, $\chi_c(1^3P_0)$,
$\chi_c(1^3P_2)$, and also three-gluon annihilation rate of
$J/{\psi}$ only the choice of larger coupling, $\alpha_s=0.191\pm
0.004$, gives rise to agreement with experiment. Even larger
$\alpha_s=0.25(2)$ provides correct number for the $\psi'$ width
 $\Gamma(\psi'\rightarrow 3g)$. Thus our
analysis shows that low-lying charmonium states have no an
universal scale for different annihilation decays  and therefore  any
ratio of their widths cannot be used to extract characteristic
strong coupling (for the discussion see \cite{16}, \cite{17}); in
particular, they are different for the $J/\psi$ and $\psi'$
three-gluon annihilation rates.

Calculated here  $\Gamma_{ee}$ for $J/{\psi},\ \psi'=\psi(3686)$,
and $\psi^{''}=\psi(3770)$ (with the mixing angle $\theta = 11^o$)
agree with experiment with accuracy $\leq 5 \%$ and this allows us
to extract some important factors from  the di-electron widths.
The essential fact is that  correct ratio of the branching
fractions, $R_{LH}=\frac{\mathcal {B}(J/\psi\rightarrow
\textrm{light\ hadrons})} {\mathcal{B}(\psi'\rightarrow
\textrm{light\ hadrons})}=0.24,$ appears to be two times larger
then in the "$12 \%$ rule" mostly because  different $\alpha_s$
describe corresponding annihilation rates.

The  unclear situation still remains with two-photon width of
$\eta_c'$, because its value can depend on possible influence of
virtual decay channel $D\bar{D}^*$ and possibly other channels
\cite{18}. In closed-channel approximation
  $\Gamma_{\gamma \gamma}(\eta_c')= 3$ keV is obtained if
the same $\alpha_s=0.24$,  which provides correct  number
for $\Gamma(\eta_c'\rightarrow gg)=11.2$ MeV, is taken. This  two-photon
width is larger than in the  CLEO experiment \cite{19}, where
under assumption (unconfirmed ) that
$\mathcal{B}(\eta_c\rightarrow KK\pi) =
\mathcal{B}(\eta_c'\rightarrow K K \pi),$ the value
$\Gamma_{\gamma\gamma}(\eta_c')=1.3\pm 0.6$ keV has been reported.
However, if via the $D\bar{D}^*$ channel the mixing of the
$3^1S_0$ and $2^1S_0$ states occurs, then even with small $3^1S_0$
contribution  to the w.f. of $\eta_c'$ ($~4\%$ to the norm) its
two-photon width is becoming essentially smaller,
$\Gamma_{\gamma\gamma}\leq 1.9$ keV.

\section{Vector and Pseudoscalar Decay Constants}

The decay constants $f_V$ and $f_P$ are calculated here with the
use of the analytic expressions, derived  in \cite{6}. To obtain
these expressions the functional integral representation for the
correlator of the currents (in V and P channels) is used and on
the final stage  this correlator is expanded in the complete set
of the eigenfunctions (e.f.) of the relativistic string
Hamiltonian (RSH) \cite{20}, \cite{21}. As the first step  we use
here this RSH to calculate charmonium spectrum and define
relativistic corrections to the decay constants:
\begin{eqnarray}\label{1}
    f^2_V(nS)&=&12\frac{\left|\varphi_n(0)\right|^2}{M_V(nS)}\ \xi_V=
\frac{3}{\pi}\frac{\left|R_n(0)\right|^2}{M_V(nS)}\ \xi_V,\\
\label{2}
    f^2_P(nS)&=&12\frac{\left|\varphi_n(0)\right|^2}{M_P(nS)}\ \xi_P=
\frac{3}{\pi}\frac{\left|R_n(0)\right|^2}{M_P(nS)}\ \xi_P.
\end{eqnarray}
Here the relativistic factors $\xi_P$, $\xi_V$, refereing to the
P and V channels, are different and given by the expressions:
\begin{equation}\label{3}
    \xi_V=\frac{m^2+\omega^2+\frac{1}{3}<\vec{p}^2>}{2\omega^2},\quad
    \xi_P=\frac{m^2+\omega^2-<\vec{p}^2>}{2\omega^2}.
\end{equation}
The values of $\omega$ and the wave functions (w.f.) at the origin are
given in Appendix. In (\ref{1}) and (\ref{2}) $R_n(0)$
(n=1,2) refers to the physical radial w.f. at the origin for  $J/\psi$,
$\psi'=\psi(3686)$, and $R_D(0)$ is the w.f. of $\psi^{''}(3770)$,
 i.e. the S-D mixing (with the mixing angle $\theta=11^o$) is
taken into account. In Appendix for  pure $2S$ and $1D$ states their w.f.
at the origin are  denoted  as $\tilde R_2(0), \tilde R_D(0)$.

The characteristic feature of RSH is that it contains a minimal
number of fundamental parameters: the string tension $\sigma$,
$\Lambda_{QCD}$ for $n_f=4$, and the pole quark mass $m_c$. The
string tension is taken from the analysis of the Regge trajectories
of light-light mesons \cite{21} and the spectra of heavy-light
mesons \cite{6}, where the preferable value is $\sigma=0.180$
GeV$^2$. The pole mass of $c$ quark is now known  with rather good
accuracy, $m_c=(1.40\pm 0.05)$ GeV \cite{22} and for $n_f=4$ the
QCD constant $\Lambda_{\overline{MS}}= 255(5)$ MeV  is taken here.

The RSH for a meson can be presented as in \cite{23}:
\begin{equation}\label{4}
     H_0=\frac{\mathbf{p}^2+m_c^2}{\omega}+\omega+V_0(r).
\end{equation}
This Hamiltonian $H_0$ is unperturbed part of general Hamiltonian,
\begin{equation}\label{5}
     H=H_0+\Delta H,
\end{equation}
where $\Delta H=V_{SD}(r)+\Delta V_{str}+V_{SE}$ includes the
spin-depended part $V_{SD}=V_{SS}+V_{LS}+V_{T}$, the string correction
$V_{str}$, and the self-energy term $V_{SE}$, which are considered
as a perturbation. Notice that  in heavy quarkonia the string correction
(as well as the self-energy term)
is always small, $\mid V_{str}\mid \leq 5$ MeV, and can be neglected.
The Hamiltonian $H_0$ has an advantage as compared to the
spinless Salpeter equation (SSE), since its w.f.  at the
origin  for $L=0$ is a regular function while the $S$-wave w.f. of SSE
diverges at small r \cite{24} and has to be  regularized.

In einbein approximation the spin-averaged mass $\overline{M}_{nL}$ can
be presented as:
\begin{equation}\label{6}
    \overline{M}_{nL}=\omega_{nL}+\frac{m_c^2}{\omega_{nL}}+
E_{nL}(\omega_c)+\Delta_{SE},
\end{equation}
where the e.v. $E_{nL}$ are the solutions of the so-called
einbein  equation:
\begin{equation}\label{7}
    \left[\frac{\vec{p}^2}{\omega_{nL}}+V_0(r)
    \right]\varphi_{nL}(r)=E_{nL}\varphi_{nL}.
\end{equation}

Notice that the mass formula (\ref{6}) does not contain an
arbitrary (fitting) constant $C_0$. It is essential that  in
einbein approximation for a given state the values,
$\overline{M}_{nL}$ and $\omega_{nL}$, are defined from the
extremum condition:
 $\frac{\partial \overline{M}_{nL}}{\partial
\widetilde{\omega}}=0$, which provides  the accuracy $\sim 5-7\% $
\cite{23}. For the mass  (\ref{6}) this extremum condition gives rise
to the relation:
\begin{equation}\label{8}
    \omega^2_{nL}=m^2_c-\frac{\partial E_{nL}}{\partial \omega_{nL}}.
\end{equation} Then for a given nL state the dynamical mass $\omega_{nL}$
and $\overline{M}_{nL}$ are calculated. Notice that $\omega_{nL}$,
being the kinetic energy of a quark, plays the role of
 the constituent quark mass which  slightly differs for the states
with different quantum numbers.

\begin{widetext}
\center
\begin{table}[h] \caption{The charmonium spin-averaged masses
$\overline{M}_{nL}$(in GeV) with the parameters ($m=1.41$ GeV,
$\sigma_0=0.18$ GeV$^2$, $\Delta_{SE}=-22$ MeV,
$\Lambda_B(n_f=4)=360$ MeV, $M_B=1.0$ GeV)$^*$.} \center
\begin{tabular}{ccccc}
  \hline  \hline
  \quad\quad State \quad\quad & \quad\quad$\omega$\quad\quad &\quad\quad$\overline{M}_{nL}$\quad\quad  &
  \quad\quad   $\overline{M}_{nL}$\quad\quad &\quad\quad Experiment\quad\quad\\
        &                                                    &                     &       $\sigma=\sigma (r)$  &   $M_{cog}$ \\
 \hline
  1$S$    & 1.592                            & 3.068                &3.066
& 3.068 \\
  2$S$    & 1.652                           & 3.678                & 3.670
& 3.674(1) \\
  1$P$    & 1.618                          & 3.488                &  3.484
& 3.525(1) \\

    \multicolumn{5}{c}{Above $D\overline{D}$ threshold}\\
  1$D$    & 1.650                            & 3.787                &
3.779                       & 3.771(2) $\uparrow$ \\
  2$P$  & 1.683                           & 3.954                &
3.940             & $\sim$ 3.930  \\
  3$S$    & 1.712                           & 4.116                &
4.093                 & 4.040 $\downarrow$ \\
  2$D$  & 1.715                            & 4.189                &
4.165                      & 4.153(3) $\uparrow$  \\
  3$P$  & 1.742                            & 4.338                &
4.299                      &  \\
  4$S$  & 1.772                            & 4.482                  &
4.424                      & 4.421(4) $\downarrow$\\
  3$D$  & 1.772                            & 4.537                  &
4.475                      &  \\
  5$S$  & 1.826                            & 4.806                  &
4.707                      & $\sim$4.664(16) \\
\hline  \hline
\multicolumn{5}{c}{}\\
\multicolumn{5}{c}{$^*$ The symbols ($\uparrow$, $\downarrow$,
$\sim$) mean that not all members of a multiplet are }\\
\multicolumn{5}{c}{measured and therefore the center of gravity
cannot be accurately }\\
\multicolumn{5}{c}{defined.}

\end{tabular}
\end{table}
 \end{widetext}

The masses given in Table 1 are calculated with
the static potential which  contains perturbative gluon-exchange
(GE) term and nonperturbative confining term:
\begin{equation}\label{9}
    V_0(r)=-\frac{4}{3}\frac{\alpha_{B}(r)}{r}+\sigma r,
\end{equation}
where the vector coupling in coordinate space $\alpha_{B}(r)$ is
defined as in \cite{25}:
\begin{eqnarray}\nonumber
&& \alpha_B(r)=\frac{2}{\pi}\int\limits_0^\infty
dq\frac{\sin(qr)}{q}\alpha_B(q),\\
 \label{10} &&
    \alpha_B(q)=\frac{4\pi}{\beta_0t_B}\left(1-\frac{\beta_1}{\beta_0^2}\frac{\ln
    t_B}{t_B}\right)
\end{eqnarray}
with $t_B=\ln\frac{q^2+M_B^2}{\Lambda_B^2}$. Here
$M_B(\sigma,\Lambda_B)=(1.00\pm0.05)$ GeV is so-called background
mass \cite{25}, and $\Lambda_B(n_f)$ can be expressed through
$\Lambda_{\overline{MS}}$: in 2-loop approximation
$\Lambda_B(n_f=4)=0.360(10)$ MeV corresponds to the
$\Lambda_{\overline{MS}}=0.254(7)$ GeV and in this case
the freezing value $\alpha_{crit}=0.547$.

Although here we consider  low-lying  states, to represent gross
features of the  charmonium spectrum and the position of higher
levels we take into account flattening of linear confining
potential. This phenomenon  occurs due to creation of virtual
light-quark pairs \cite{21} and it is becoming essential for
higher levels, in particular, for the mass of the $5^3S_1$ state.
The flattening of linear potential is defined  by the analytic
function for which the form and parameters  are taken  just the
same as in \cite{21}, where they have been extracted  from the
light meson radial Regge trajectories. The origin of flattening
comes from  the virtual $q\bar{q}$ pairs creation on  the surface
inside the Wilson loop $\langle W(C)\rangle$, having  large size,
and due to these virtual loops  the string tension (as well as the
surface) is becoming smaller and dependent on the $Q\bar{Q}$
separations r. This potential provides a correlated mass shift
down of all radial excitations with $n\geq 3$.

From Table 1 one can also see that in the potential with
$\sigma(r)r$ the masses M(3S),M(4S), M(5S) are shifted down by
$\sim20$ MeV, $\sim 60 $ MeV, and $\sim 100$ MeV, respectively,
and turn out to be close to the experimental values with the
exception of the $\psi(4040)$, which is strongly affected by the
nearby $S$ wave threshold. Since the hyperfine splitting of the $5^3S_1$
level is small, $\leq 6$ MeV \cite{13},  its mass
practically coincides with the centroid mass, $M(5S)=4.70$ GeV,
calculated here, and lies close to the mass of the Belle resonance
Y(4660) \cite{4}.  From Table 1 one can see that with exception of the
$1P$ state all other masses are in good agreement with experiment, even
for the states above the $D\bar{D}$ threshold. The reason, why only for
the $1P$ states the centroid mass has smaller value (in einbein
approximation) needs an additional analysis.

The V and P decay constants, calculated with the use of  (\ref{1})
and (\ref{2}), are given  in Tables 2 and 3 together  with those
from some other papers.

 \begin{widetext}
 \center
\begin{table}[h]
\caption{The decay constants $f_V$ (in MeV) of the $J/\psi$,
$\psi'=\psi(3686)$, $\psi''=\psi(3770)$ mesons.} \center
\begin{tabular}{ccccccc}
  \hline  \hline
  State    & This paper & BL \cite{9} & EFG \cite{11}  &
 Wang\cite{12} &
\multicolumn{2}{c}{Experiment} \\\cline{6-7}
           &            &    (Rel)&     &       &  $\beta_V=1.0$   & $\beta_V=0.72 $  \\
           &            &         &     &       &  $(\alpha_s=0)$
& $(\alpha_s=0.165)$  \\
  \hline
  $J/\psi$ & 483        &  545 & 551 & 459(28)  & 415(6) & 490(7)\\

  $\psi'$  & 357        &  371 & 401 & 364(24)  & 302(4) & 356(4)\\

  $\psi''$ & 115        &  318 &     &         & 96(5)  & 113(6)\\
    \hline  \hline
\end{tabular}
\end{table}

\begin{table}[h]
\caption{The decay constants  $f_P$ (in MeV) of the $\eta_c$, $\eta_c'$,
$\eta_{c2}$ mesons.} \center
\begin{tabular}{cccccc}
  \hline  \hline
  State & This paper & BL \cite{9}   & LP\cite{10} &
\multicolumn{2}{c}{Experiment} \\\cline{5-6}
        &            &     (Rel)  &         & $\beta_P=1.0$   & $\alpha_s(\eta_c)=0.195$ \\
        &            &            &          & $(\alpha_s=0)$    & $\alpha_s(\eta_c')=0.25$ \\
  \hline
  $\eta_c$        & 453 & 493 &   480 & 404(57) & 454(64) \\

  $\eta_c'$       & $336^{a)}, \ \ 267^{b)}$  & 260   & 303 & 189(40) & 213(47)\\

  $\eta_{c2}$      & 41 &     &    &         & \\
  \hline  \hline
\multicolumn{6}{c}{}\\
 \multicolumn{6}{c}{$^{a)}$ The influence of virtual decay channels is
neglected.}\\
 \multicolumn{6}{c}{$^{b)}$ Mixing of $2^1S_0$ and $3^1S_0$ states ($\theta=11^o$) is taken into account.}
\end{tabular}

\end{table}

 \end{widetext}

 It is worth pointing out that the "experimental"
$f_V$ and $f_P$, extracted from the experimental di-electron and
two- photon widths, depend on chosen values of  radiative
corrections, $\beta_V$ and $\beta_P$. Therefore in Tables 1 and 2
we give two variants of "experimental" decay constants, which
correspond to $\beta_V= 0.72\ (\alpha_s=0.165)$ and 1.0, and
$\beta_P=0.79\ (\alpha_s=0.195)$ and 1.0 $(\alpha_s=0)$.
It is also important that $\eta_c'(2^1S_0)$ and $\eta_{c2}(1^1D_2)$
cannot be mixed and therefore their w.f. at the origin are
defined by the w.f. $\tilde R_{2S}(0)$ and $\tilde R_D(0)$
for pure $2S$ and $1D$ states (see Appendix).
From Table 3 one can see that for $\beta_V=0.72$ our decay
constants $f_V$ are in good agreement with experiment. Note that  for
$\psi''$  calculated  constant $f_V$  is almost three
times smaller than in \cite{9}.

\section{Di-electron widths}
The leptonic width of a vector state in heavy quarkonia is
expressed via the decay  constant  \cite{6},\cite{26}:
\begin{eqnarray}\label{11}
   && \Gamma_{ee}(V)=\frac{4 \pi e_c^2\alpha^2}{3 M_V}f_V^2
   \beta_V,\\
\label{12}
   && \beta_V=1-\frac{16}{3\pi}\alpha_s.
\end{eqnarray}

Best description of  di-electron widths   is obtained here taking
in (\ref{11}) $\alpha_s=0.165$, or  $\beta_V=0.72,$  for which
$\Gamma_{ee}(J/\psi)=5.41$ keV, $\Gamma_{ee}(\psi')=2.47$ keV, and
$\Gamma_{ee}(\psi^{''})=0.248$ keV are obtained (see Table 4). In
our calculations of the di-electron widths of $\psi'(3686)$ and
$\psi''(3770)$ the S-D mixing  with the mixing angle $\theta=11^0$
is taken into account (In Appendix we give their physical w.f. at
the origin and also several m.e.). In Table 4 calculated widths
are presented together with some  theoretical predictions and
experimental data.

Since the coupling in (\ref{12}) appears  to be the same for
$J/\psi, \ \psi',\ \psi''$, it is of interest to compare  their
ratios (where the radiative corrections are cancelled):
 $R_{\psi'}=\Gamma_{ee}(\psi')/\Gamma_{ee}(J/\psi)$,
$R_{\psi''}=\Gamma_{ee}(\psi'')/\Gamma_{ee}(J/\psi)$   with
experimental numbers which turn out to be very close to each other:
\begin{equation}\label{13}
\begin{array}{cc}
  (R_{\psi'})_{th}=0.46, & (R_{\psi'})_{exp}=0.45\pm 0.02, \\
  (R_{\psi''})_{th}= 0.046, & (R_{\psi'})_{exp}=0.044\pm 0.03. \\
\end{array}
\end{equation}

Having such  agreement with experiment we expect that our w.f. at
the origin are defined with good accuracy, in particular, the  following
numbers are obtained for the factors :
\begin{equation}\label{14}
\begin{array}{cc}
\frac{\left|R_{J/\psi}(0)\right|^2}{M^2_{J/\psi}}=0.085 GeV, &
\frac{\left|R_{\psi'}(0)\right|^2}{M^2_{\psi'}}=0.040 GeV,
\end{array}
\end{equation}

\begin{equation}\label{15}
\begin{array}{cc}
\frac{\left|R_{\psi''}(0)\right|^2}{M^2_{\psi''}}=0.0040 GeV, &
\frac{\left|R'_{p}(0)\right|^2}{M_{cog}^2\omega_P^2}=0.0027 GeV, \\
\end{array}
\end{equation}
 We estimate the accuracy for these factors as $\leq 10 \%$  and later
use them to define  three- and two-gluon
annihilation rates for $J/\psi,\ \psi',\ \eta_c,\ \eta'_c $.

\begin{table}[h]
\caption{The di-electron widths (in keV) (with $\alpha_s=0.165$
and the mixing angle $\theta=11^o).$} \center
\begin{tabular}{ccccc}
  \hline  \hline
  State & This paper & BL\cite{27} & EFG \cite{11}  & Experiment \\
  \hline
  $J/\psi$  & 5.41  & 12.13 & 5.4  & $5.55\pm 0.16$ \\
  $\psi' $  & 2.47  & 5.03  & 2.4  & $2.48\pm 0.06$ \\
  $\psi''$  & 0.248 & 0.056 &      & $0.242^{+0.027}_{-0.024}$\\
  \hline  \hline
\end{tabular}
\end{table}
Notice that  relativistic corrections decrease the decay constants
$f_V$ and $f_P$ and  provide  better agreement with experiment.
However, for hadronic decays this type of relativistic corrections
is not still calculated in FCM, and therefore the accuracy of
calculated hadronic widths  is worse than for the decay constants
$f_V$ and $f_P$. Concluding this Section, we would like to
underline that our calculations give the ratio of the branching
fractions for the $e^+ e^-$ annihilation:

\begin{equation}\label{16}
\begin{array}{c}
                R_{ee}=\frac{\mathcal{B}_{ee}(\psi')}{\mathcal{B}_{ee}(J/\psi)}=\frac{\Gamma_{tot}(J/\psi)}{\Gamma_{tot}(\psi')}
  \frac{|R_{\psi'}(0)|^2}{|R_{J/\psi}(0)|^2}\frac{M^2(J/\psi)}
  {M^2(\psi')} \frac{\xi_V(2S)}{\xi_V(1S)}= \\
 =(12.6\pm 0.03) \%,
\end{array}
\end{equation}
which is in good agreement with experimental number, or the
so-called "$12 \%$ rule", $R_{ee}(exp)=(12.3\pm 0.03)\%$.

\section{Two-photon widths of $\eta_c,\eta_c',\chi_c(1^3P_0),
\chi_c(1^3P_2)$}

 The two-photon widths of $\ \eta_c(1S),\
\eta_c'(2S)$ can be expressed via the decay constants $f_P$:
\begin{eqnarray}\label{17}
    &&\Gamma_{\gamma\gamma}(P)=\frac{4\pi
    \alpha^2e_c^4}{M_P}f_P^2\beta_P,\\
    \label{18}
    &&\beta_P=1-\frac{20-\pi^2}{3\pi}\alpha_s.
\end{eqnarray}
The two-photon widths of the scalar  $\chi_{c0}$ and tensor
$\chi_{c2}$ mesons can be derived in FCM the same manner as it has
been done for the V, P decay constants in \cite{6} with the
following result:

\begin{equation}\label{19}
   f_{S(T)}^2=\frac{3|R_P'(0)|^2}{\pi M_{S(T)}\omega_P^2}.
\end{equation}
Here in (\ref{19})  instead of the quark mass $m_Q$ (in
nonrelativistic limit) the kinetic energy $\omega_P$ enters. Then
with the QCD corrections the two-photon widths of the $P-$wave
mesons are defined as \cite{7}:
\begin{eqnarray}\label{20}
    \Gamma_{\gamma\gamma}(\chi_{c_0})&=&\frac{108\alpha^2e_c^4
\left|R'_P(0)\right|^2}{M_S^2\omega_P^2}\left(1-\left(\frac{28-3\pi^2}{9}
\right)\frac{\alpha_s}{\pi}\right), \\
\label{21}
    \Gamma_{\gamma\gamma}(\chi_{c_2})&=&\frac{144\alpha^2e_c^4
\left|R'_P(0)\right|^2}{5M_T^2\omega_P^2}\left(1-\frac{16}{3}\frac{\alpha_s}
{\pi}\right).
\end{eqnarray}

In our analysis we use $\alpha_s=0.195$ for $\eta_c$, $\chi_{c0}$,
and $\chi_{c2}$ to obtain good numbers for the following ratios
of two-photon widths (in which relativistic factors
 $\xi_P(\eta_c)=0.785, \ \xi_P(\eta_c') =\xi_P(\eta_{c2})=0.73$ are
used, see Appendix):

\begin{equation}\label{22}
\begin{array}{cc}
 \left(\frac{\Gamma_{\gamma\gamma}(\chi_{c_0})}{\Gamma_{\gamma\gamma}
(\eta_c)}\right)_{th}=0.458 ,
   &\left(\frac{\Gamma_{\gamma\gamma}(\chi_{c_0})}{\Gamma_
{\gamma\gamma}(\eta_c)}\right)_{exp}= 0.41\pm 0.02. \\\end{array}
\end{equation}
For $\chi_{c0}$ the width  is by 10 \% larger than the experimental
one, while  for $\chi_{c2}$ the absolute value and the ratios:

\begin{equation}\label{23}
\begin{array}{cc}
 \left(\frac{\Gamma_{\gamma\gamma}(\chi_{c_2})}
{\Gamma_{\gamma\gamma}(\eta_c)}\right)_{th}=0.075, &
 \left(\frac{\Gamma_{\gamma\gamma}(\chi_{c_2})}{\Gamma_
{\gamma\gamma}(\eta_c)}\right)_{exp}=0.076\pm 0.002 \\
\end{array}
\end{equation}
coincide with the  experimental values. Different two-photon widths are
given in Table 5.

\begin{table}[h]
\caption{The  two-photon widths (in keV) with $\alpha_s=0.195$ for
$\eta_c$,  $\chi_{c0},$ $\chi_{c2}$ and with $\alpha_s=0.24$ for
$\eta'_c$,  $\eta_{c2}$. }  \center
\begin{tabular}{cccccc}
  \hline  \hline
  State & This paper & BL \cite{9} & EFG \cite{11}  & KLW \cite{12} &
Experiment \\
  \hline
  $\eta_c$           &  7.22 & 4.18  & 5.5 & 7.14(95) & $7.1\pm2.7$
\cite{22} \\
  $\eta_c' $         &  3.0$^{a)}$,  \ 1.9$^{b)}$& 2.59  & 1.8 & 4.44(48) & $1.3\pm0.6$
\cite{18}
\\
  $\eta_{c2}$          & 0.042  & 1.21  & -    &          &  \\
  $\chi_{c0}(1^3P_0)$ & 3.31  & 3.28  & 2.9  & 3.78         & 2.90 $\pm$
0.43
\\
  $\chi_{c2}(1^3P_2)$ & 0.54  &  -    & 0.52 &          & 0.53 $\pm$ 0.06
\\
\hline  \hline
\multicolumn{6}{c}{}\\
 \multicolumn{6}{c}{$^{a)}$ See the footnote $^{a)}$ to Table 3.}\\
 \multicolumn{6}{c}{$^{b)}$ See the footnote $^{b)}$ to Table 3.}
\end{tabular}
\end{table}

We would like to stress here that calculated two-photon width of
$\eta_c'$ (with $\alpha_s=0.24$) is larger than in the CLEO
experiment \cite{19}, nevertheless with the same $\alpha_s$ we
have obtained hadronic width $\Gamma(\eta_c'\rightarrow gg)=11.0$
MeV in agreement with $\Gamma_{tot}=14\pm 7$ MeV \cite{22}.

One cannot exclude that the w.f. at the origin of $\eta_c'$ is
affected by the virtual $D\bar{D}^*$ decay channel which lies only
by 130 MeV higher than $\eta_c'$. Then  via this channel the w.f.
of $\eta_c'$ can be mixed  with $3^1S_0$ state (which is now often
identified with $X(3940)$ \cite{29}). For example, the 20\%
admixture in the w.f. of $\eta_c'$ (or the 4\% contribution to the
norm) gives rise to the two-photon
$\Gamma_{\gamma\gamma}(\eta_c')=1.9$ keV.

Recently  two-photon width of $\eta_c$ has been calculated
in lattice QCD \cite{30}, \cite{31}. Such calculations from first
principles are very important for the theory, however, in \cite{30}
 rather small number, $\Gamma_{\gamma\gamma}(\eta_c)=
2.65(26)(80)(53)$ keV and corresponding $f_P(\eta_c)=373$ MeV are
obtained in quenched approximation. Meanwhile in \cite{31} the
calculations  in lattice QCD with exact chiral symmetry, where
heavy quarks  are treated  as the Dirac fermions, the  larger
decay constant, $f_P(\eta_c)=438(11)$ MeV, is obtained and this
number is in agreement with our result, $f_P(\eta_c)=453$ MeV.

\section{Three- and  two-gluon annihilation rates}

The QCD corrections to di-electron widths have appeared to be
$\leq 30\%$, but they are even more important for some
hadronic decays. From \cite{7}, \cite{8},
\cite{14} we know the widths of the three-gluon annihilation and the decay
into $\gamma gg$ for the vector mesons, as well as for  two-gluon
annihilation of P, S, T mesons:
\begin{eqnarray}\label{24}
    \Gamma(V\rightarrow
    ggg)&=&\frac{40(\pi^2-9)\alpha_s^3}{81\pi M_V^2}\left|R_n(0)
\right|^2\left(1-3.7\frac{\alpha_s}{\pi}\right),\\
\label{25}
    \Gamma(V\rightarrow \gamma gg)&=&\frac{32(\pi^2-9)\alpha_s^2\alpha}{9\pi M_V^2}\left|R_n(0)
\right|^2\left(1-6.7\frac{\alpha_s}{\pi}\right),\\
\label{26}
    \Gamma(P\rightarrow
gg)&=&\frac{8\alpha_s^2}{3M_P^2}\left|R_n(0)\right|^2\left(1+4.8
\frac{\alpha_s}{\pi}\right),\\
\label{27}
    \Gamma(\chi_{c_0}\rightarrow
gg)&=&\frac{24\alpha_s^2}{M_S^2\omega_P^2}\left|R'_P(0)\right|^2\left(1+9.5\frac{\alpha_s}{\pi}\right),\\
\label{28}
    \Gamma(\chi_{c_2}\rightarrow
gg)&=&\frac{32\alpha_s^2}{5M_T^2\omega_P^2}\left|R'_P(0)\right|^2\left(1-2.2\frac{\alpha_s}{\pi}\right).
\end{eqnarray}

It can be easily shown that with $\alpha_s=0.165$, as for considered above
di-electron widths,  the three-gluon annihilation width of
$J/\psi$ is smaller than in experiment being equal 42 keV. The
best fit to this annihilation rate is obtained taking
$\alpha_s=0.187$ (practiclly the same as in  two-photon
width), for which
\begin{eqnarray}
\nonumber\Gamma(J/\psi\rightarrow ggg)&=&59.5\ keV,\\
\nonumber\Gamma(J/\psi\rightarrow \gamma gg)&=&5.7\ keV.
\end{eqnarray}
 Then the sum of these annihilation widths , being equal the width
$\Gamma(J/\psi\rightarrow  \textrm{light\ hadrons})=65.2$ keV, is
in good agreement with the experimental number
$\mathcal{B}(J/\psi\rightarrow \textrm{light\ hadrons})=(69\pm
3)\%$ \cite{16}, \cite{17} or $\Gamma(J/\psi\rightarrow
\textrm{light\ hadrons})_{exp}=64 \pm 3$ keV.

On the other hand this value, $\alpha_s=0.187$, is not sufficient to
provide correct number for the three-gluon annihilation rate of
$\psi'(3686)$ and in this case the best fit is obtained taking
$\alpha_s=0.238$:
\begin{eqnarray}
    \nonumber\Gamma(\psi'\rightarrow ggg)&=&52.7\ keV,\\
    \nonumber\Gamma(\psi'\rightarrow \gamma gg)&=&3.5\ keV,
\end{eqnarray}
with their  sum $\Gamma(\psi'\rightarrow \textrm{light\
hadrons})=56.2$ keV, in good agreement with the experimental
branching fraction $\mathcal{B}(\psi'\rightarrow \textrm{light\
hadrons}) =(16.9\pm 3)\%$ \cite{16}, \cite{17}, or
$\Gamma(\psi'\rightarrow \textrm{light\ hadrons})_{exp}=57 \pm 10$
keV. This fact means that  the ratio

\begin{equation}\label{29}
\begin{array}{c}
  R_{LH}= \frac{\mathcal{B}(\psi'\rightarrow \textrm{light\ hadrons})}{\mathcal{B} (J/\psi\rightarrow  \textrm{light\ hadrons})}= \\
 =\frac{\Gamma_{tot}(J/\psi)}
{\Gamma_{tot}(\psi')}\frac{\Gamma(\psi'\rightarrow \textrm{light\
hadrons})} {\Gamma(J/\psi\rightarrow  \textrm{light\
hadrons})}=0.24
\end{array}
\end{equation}
is in agreement with experiment and two times larger than the
ratio $R_{ee}$ (\ref{16}) mostly because different $\alpha_s$ are
taken for $J/\psi$ and $\psi'$.

We would like to notice also that in our calculations the ratio,

$$R_{th}=\frac{\Gamma(J/\psi\rightarrow \textrm{light\ hadrons})}{\Gamma_{ee}(J/\psi)}=11.0\pm 0.05$$
 agrees with experimental number,  $R_{exp}=11.6\pm 0.03$, and this fact
justifies  our choice of $\alpha_s=0.187$ in radiative correction
for $J/\psi$.

\section{Conclusions}

We have shown that di-electron widths of $J/\psi,\ \psi',\
\psi^{''}$ describe experimental data with the accuracy better
$5\%$, if  the same $\alpha_s=0.165$ is taken in the QCD radiative
corrections. This fact can be considered as a test of the method
used here.

For $\eta_c(1S),\ \chi_{c0},\ \chi_{c2}$ the larger effective
coupling, $\alpha_s=0.195$, is needed to fit experimental numbers
for two-photon widths. Also close value of $\alpha_s=0.187$
provides good description of the annihilation widths for the
$J/\psi\rightarrow ggg$ and $J/\psi\rightarrow \gamma gg$
processes, so that $\mathcal{B}(J/\psi\rightarrow light \
hadrons)=70\%$ is obtained (its value in experiment is $(69\pm
3)\%$ \cite{16}, \cite{17}).

However, to describe experimental branching fraction,
$\mathcal{B}(\psi'\rightarrow \textrm{light\ hadrons})=16.7\pm
3.0\%$ \cite{16}, \cite{17}, larger $\alpha_s=0.238$ is needed,
for which we obtain $\Gamma(\psi'\rightarrow ggg)=52.7$ keV and
$\Gamma(\psi'\rightarrow \gamma gg)=3.5$ keV, so that their sum is
equal $\Gamma(\psi'\rightarrow light \ hadrons)=56.2$ keV, which
corresponds to experimental branching fraction. Then the ratio  of
the branching fractions:
         $$R_{LH}=\frac{\mathcal{B}(\psi'\rightarrow \textrm{light\ hadrons})}
          {\mathcal{B}(J/\psi \rightarrow \textrm{light\ hadrons})}=0.24 \pm
          0.01,$$
 appears to be in good agreement with experimental number,
$R_{LH}(exp)=0.24\pm 0.04$, being two times larger than the ratio
$R_{ee}$ (\ref{16}).

Nevertheless for the same $\alpha_s=0.24$ calculated here
two-photon width of $\eta_c'$,
$\Gamma_{\gamma\gamma}(\eta_c')=3.0$ keV, turns out to be larger
than in the CLEO experiment \cite{18}. From our point of view the
problem of this width can be solved taking into account the
influence of the $D\bar{D}^*$ decay channel.

Thus  our analysis shows that  there is no an universal effective
strong coupling  which allows to describe different annihilation
processes for all low-lying charmonium states. Such universal
description is possible only for di-electron widths.
\section*{Acknowledgements}
This work is supported by the Grant NSh-843.2006.2~.

\appendix
\section*{Appendix: The matrix elements and wave function at the origin of low-lying
charmonium states}

Here we  use  RSH to calculate the m.e., like $\omega(nL),\
<\vec{p}^2>$, and also the w. f. at the origin. For pure
 $2S$ state we denote its w.f. as $\tilde R_{2S}(0)=0.767$ GeV$^{3/2}$; for
the $P$-wave states $\tilde R_P(0)=R_P'(0)/\omega_P=0.183$
GeV$^{3/2}$ with $R_P'(0)=0.297$ GeV$^{5/2}$, and for the $1D$
state  $\tilde R_D(0)= \frac{5
R_D''(0)}{2\sqrt{2}\omega_D^2}=0.0942$ GeV$^{3/2}$ with
$R_D''(0)=0.145$ GeV$^{7/2}$. The values of $\omega(nL)$ are given
in Table 6.

The physical w.f. of $\psi'$ and $\psi''$, calculated here,
take into account  the mixing angle $\theta=11^o$, and these w.f. at the
origin are given in Table 6 (all needed parameters are given in
Section 2).

 For $\psi'$ and $\psi^{''}$   the S-D mixing  can occur due to
tensor forces and coupling to  open $D \bar{D}$ channel. Then
the w. f. at the origin of $\psi'$ and
$\psi^{''}$ are defined as in \cite{28}:
\begin{eqnarray}\label{30}
&&R_{\psi'}(0)=\cos\theta \tilde{R}_{2S}(0)-
\frac{5}{2\sqrt{2}\omega^2}\sin\theta R''_D(0),\\ \label{31}
&&R_{\psi''}(0)=\sin\theta \tilde{R}_{2S}(0)+
\frac{5}{2\sqrt{2}\omega^2}\cos\theta R''_D(0).
\end{eqnarray}

Our calculations give $R_P(0)=\frac{|R_P'(0)|^2}{\omega_P}=0.183$
GeV$^{3/2}$; the $\psi''$ w.f. at the origin, $R_D(0)=0.238$
GeV$^{3/2}$ ($\omega(1D)=1.65$ GeV), and for the $\psi'$ meson
$R_{\psi'}(0)=0.735$ (see Table 6).

\begin{table}[h]
\caption{The dynamical masses $\omega_{nL}$,  the radial wave
functions at the origin for $J/\psi$ ,$\psi'$, $\psi^{''}$, and
$\chi_{cJ}$ (in GeV).} \center
\begin{tabular}{cccccc}
  \hline  \hline
   State     &   $\omega_{nL}$, GeV  & $R_{nL}(0)$
  &$<\vec{p}^2>$  &$\xi_V$& $\xi_P$\\
  \hline
  1S   & 1.59 & 0.905 & 0.541 & 0.929 & 0.785\\
  2S   & 1.65 & 0.735 & 0.722 & 0.910 & 0.733\\
  1P   & 1.62 & 0.183 & 0.619 & 1.0   & 1.0 \\
  1D   & 1.65 & 0.238 & 0.721 & 0.911 & 0.733\\
  \hline  \hline

\end{tabular}
\end{table}


\begin{thebibliography}{1}

\bibitem{1}
 (BELLE Collaboration) S.~K.~Choi et al., Phys.\ Rev.\ Lett.\ {\bf
 91}, 262001 (2003); (CDF Collaboraton) D.~Acosta et al., Phys.\
 Rev.\ Lett.\ {\bf 93}, 072001 (2004); (DO Collaboration)
 V.~M.~Abazov et al. Phys.\ Rev.\ Lett.\ {\bf 93}, 162002 (2004);
 (BaBar Collaboration) B.~Aubert et al., Phys.\ Rev.\ D {\bf 91},
 071103 (2005).

\bibitem{2}
 (BaBar Collaboration) B. Aubert et al., Phys.\ Rev.\ Lett.\ {\bf
 95}, 142001 (2005); Phys.\ Rev.\ D {\bf 74}, 091103 (R); (CLEO
 Collaboration) T.~E.~Coan et al., Phys.\ Rev.\ Lett.\ {\bf 96},
 162003 (2006); (BELLE Collaboration) K.~Abe et al., Phys.\ Rev.\
 Lett.\ {\bf 96}, 162003 (2006).

\bibitem{3}
 (BaBar Collaboration) B.~Aubert et al., Phys.\ Rev.\ Lett.\ {\bf
 98}, 212001 (2007).

\bibitem{4}
 (BELLE Collaboration) X.~L.~Wang  et al., Phys.\ Rev.\ Lett.\ {\bf
 99}, 142002 (2007), arXiv:hep-ex/0707.3699v1.

\bibitem{5}
 A.~M.~Badalian, B.~L.~Ioffe,and  A.~V.~Smilga, Nucl. Phys. B {\bf 281},
85 (1987)


\bibitem{6}
  A.~M.~Badalian, B.~L.~G.~Bakker, and Yu.~A.~Simonov,
    Phys.\ Rev.\  D {\bf 75}, 116001 (2007)
  [arXiv:hep-ph/0702157].




\bibitem{7}
  R.~Barbieri, M.~Caffo, R.~Gatto, and E.~Remiddi,
    Nucl.\ Phys.\  B {\bf 192}, 61 (1981);
  R.~Barbieri, E.~d'Emilio, G.~Curci, and E.~Remiddi,
   Nucl.\ Phys.\  B {\bf 154}, 535 (1979).

\bibitem{8}
  W.~Kwong, P.~B.~Mackenzie, R.~Rosenfeld, and J.~L.~Rosner,
    Phys.\ Rev.\  D {\bf 37}, 3210 (1988);  S.~J.~Brodsky, G.~P.~Lepage
 and P.~B.~Mackenzie,   Phys.\ Rev.\  D {\bf 28}, 228 (1983).


\bibitem{9}
  O.~Lakhina and E.~S.~Swanson,  Phys.\ Rev.\  D {\bf 74}, 014012 (2006).

\bibitem{10}
  J.~P.~Lansberg and T.~N.~Pham,
  Phys.\ Rev.\  D {\bf 74}, 034001 (2006): T.~ N.~ Pham, arXiv: 0710.2846;
  N.~Fabiano, G.~Pancheri,  Eur.\ Phys.\ J. C {\bf 25}, 42 (2002).


\bibitem{11}
  D.~Ebert, R.~N.~Faustov, and V.~O.~Galkin,
  Mod.\ Phys.\ Lett.\  A {\bf 18}, 601 (2003).
\bibitem{12}
 G.L.Wang, Phys.\ Lett.\ B {\bf 633}, 492 (2006);
 C.S.Kim, T.Lee, and G.L.Wang, Phys.\ Lett.\ B {\bf 606}, 323
(2005).

\bibitem{13}
  A.~M.~Badalian and B.~L.~G.~Bakker,
  Phys.\ Lett.\  B {\bf 646}, 29 (2007);
  A.~M.~Badalian, A.~I.~Veselov, and B.~L.~G.~Bakker,
  J.\ Phys.\ G {\bf 31}, 417 (2005).

\bibitem{14}
 F.Yndurain "The theory of Quark and Gluon Interaction", fourth edition,
 Springer-Verlag, Berlin-Heidelberg, p.189 (2006).

\bibitem{15}
 A.~M.~Badalian, B.~L.~G.~Bakker, and I.~V.~Danilkin,
 (in preparation)

\bibitem{16}
  E.~Eichten, S.~Godfrey, H.~Mahlke, and J.~L.~Rosner,
  Status Report "Quarkonia and their Transitions",
  arXiv:hep-ph/0701208.



\bibitem{17}
 Y.~F.~Gu and X.~H.~Li, Phys.\ Rev.\  D {\bf 63}, 114002 (2001);
 M.~B.~Voloshin,  arXiv: 0711.4556; K.~K.~Seth, arXiv: hep-ex/0511062; hep-ex/0504052.
\bibitem{18}
 E.~J.~Eichten, K.~Lane, C.~Quigg , Phys.\ Rev.\ D {\bf 73}, 014014
 (2006),(Erratum-ibid. D {\bf 73}, 079903 (2006)).

\bibitem{19}
  D.~M.~Asner {\it et al.}  [CLEO Collaboration],
  Phys.\ Rev.\ Lett.\  {\bf 92}, 142001 (2004).


\bibitem{20}
  A.~Y.~Dubin, A.~B.~Kaidalov and Yu.~A.~Simonov,
  Phys.\ Atom.\ Nucl.\  {\bf 56}, 1745 (1993)
  [Yad.\ Fiz.\  {\bf 56}, 213 (1993)]
  [arXiv:hep-ph/9311344]; Phys. Lett. B {\bf 323}, 41 (1994);
  Yu.~A.~Simonov, "QCD and topics in hadron physics,''
  arXiv: hep-ph/9911237.


\bibitem{21}
  A.~M.~Badalian, B.~L.~G.~Bakker, and Yu.~A.~Simonov,
  Phys.\ Rev.\  D {\bf 66}, 034026 (2002).
  A.~M.~Badalian and B.~L.~G.~Bakker,
  Phys.\ Rev.\  D {\bf 66}, 034025 (2002).

\bibitem{22}
  W. M. Yao et al. [Particle data Group], J. Phys. G {\bf 33}, 1 (2006).


\bibitem{23}
  Yu.~S.~Kalashnikova, A.~V.~Nefediev, and Yu.~A.~Simonov,
  Phys.\ Rev.\  D {\bf 64}, 014037 (2001);   Yu.~A.~Simonov,
  Phys.\ Atom.\ Nucl.\  {\bf 67}, 553 (2004)
  [Yad.\ Fiz.\  {\bf 67}, 571 (2004)].




\bibitem{24}
  S.~Jacobs, M.~G.~Olsson, and C.~I.~Suchyta,
    Phys.\ Rev.\  D {\bf 33}, 3338 (1986)
  [Erratum-ibid.\  D {\bf 34}, 3536 (1986)].



\bibitem{25}
  A.~M.~Badalian and D.~S.~Kuzmenko,
  Phys.\ Rev.\  D {\bf 65}, 016004 (2002);  A.~M.~Badalian and
Yu.~A.~Simonov,    Phys.\ Atom.\ Nucl.\  {\bf 60}, 630 (1997)
  [Yad.\ Fiz.\  {\bf 60}, 714 (1997)].

\bibitem{26}
 R.~Van~Royen and V.~F.~Weisskopf, Nuovo Cim. {\bf 50}, 617 (1967);
 ibid 51, 583 (1967).


\bibitem{27}
   T.~Barnes, arXiv: hep-ph/0406327.

\bibitem{28}
   V. A. Novikov et al., Phys. Rep. C {\bf 41}, 1 (1978);
   J.L.Rosner, Phys.Rev. D {\bf 64}, 094002 (2001).

\bibitem{29}
   (BELLE Collaboration) K.~Abe et al., arXiv: hep-ex/0708.3812; Phys.\ Rev.\ Lett.\ {\bf 98}, 082001 (2007).
\bibitem{30}
  J. J. Dudek, and R. G. Edwards, Phys. Rev. Lett. {\bf 97}, 172001 (2006).
\bibitem{31}
(TWQCD Collaboration) T. W. Chiu, T. H. Hsieh, C.H. Huang,
and K. Ogawa, arXiv:0711.2131.
\end{thebibliography}
\end{document}